\documentclass[a4paper,11pt]{article}
\pdfoutput=1
cm \voffset = -2truecm

\usepackage{graphicx}
\usepackage{amsmath}
\usepackage{amssymb}
\usepackage{latexsym}
\usepackage[colorlinks]{hyperref}
\usepackage{color}
\usepackage{cite}
\usepackage{makeidx}
\usepackage{float}
\usepackage{multirow}
\newcommand{\bra}{\begin{array}}
\newcommand{\era}{\end{array}}
\newcommand{\beq}{\begin{equation}}
\newcommand{\eeq}{\end{equation}}
\newcommand{\beqar}{\begin{eqnarray}}
\newcommand{\eeqar}{\end{eqnarray}}

\def\BC{\bb C}
\def\_\BC{\bbi C}

\def\( {\left(}
   \def\) {\right)}
\def\[ {\left[}
\def\] {\right]}
\def\no2 {{\textstyle{n\over 2}}}

\newcommand{\om}{\omega}

\newcommand{\lam}{\lambda}
\newcommand{\si}{\sigma}

\newcommand{\be}{\beta}

\newcommand{\lga}{\longrightarrow}

\newcommand{\lb}{\label}

\begin{document}
\begin{titlepage}
\setcounter{page}{1}
\renewcommand{\thefootnote}{\fnsymbol{footnote}}

\begin{flushright}
%ucd-tpg:****.**\\
%arXiv:yymm.xxxx
\end{flushright}

\vspace{5mm}
\begin{center}

{\Large \bf {Gate-Tunable Graphene Quantum Dot and Dirac
Oscillator}}

\vspace{5mm}

{\bf Abdelhadi Belouad}$^{a}$, {\bf Ahmed Jellal\footnote{\sf ajellal@ictp.it --
a.jellal@ucd.ac.ma}}$^{a,b}$ and  {\bf Youness Zahidi}$^{a}$

\vspace{5mm}

%{$^b$\em Physics Department,  King Fahd University
%of Petroleum $\&$ Minerals,\\
%Dhahran 31261, Saudi Arabia}

{$^{a}$\em Theoretical Physics Group,  %Department of Physics,
Faculty of Sciences, Choua\"ib Doukkali University},\\
{\em PO Box 20, 24000 El Jadida, Morocco}

{$^b$\em Saudi Center for Theoretical Physics, Dhahran, Saudi Arabia}

%\vspace{30mm}

\vspace{3cm}

\begin{abstract}

We obtain the solution of the Dirac equation in (2+1) dimensions
in the presence of a constant magnetic field normal to the plane
together with a two-dimensional Dirac-oscillator potential
coupling. We study the energy spectrum of graphene quantum dot
(QD) defined by electrostatic gates. We give discussions of our
results based on different physical settings, whether the
cyclotron frequency is similar 
or larger/smaller compared to the oscillator frequency.
 This defines an effective magnetic
field that produces the effective quantized Landau levels. We
study analytically such field
in {gate-tunable} graphene QD and show that 
our structure allow us to
control the valley degeneracy. 
Finally, we compare our results with already published work and
also discuss the possible applications of such QD.

% \pacs{02.30.Gp, 02,30.Tb, 02.30.Jr}

% \maketitle

%\newpage

\end{abstract}
\end{center}

\vspace{7cm}


\end{titlepage}

%\newpage

%%%%%%%%%%%%%%%%%%%%%%%%%%%%%%%%%%%%%%%%%%%%%%%%%%
\section{Introduction}
%%%%%%%%%%%%%%%%%%%%%%%%%%%%%%%%%%%%%%%%%%%%%%%%%%%%%%

In recent years, several interest have been devoted to the
study of two-dimensional (2D) system such as quantum wells,
quantum wires, and quantum dots~\cite{GHRMB99,PBSKS00,WHB03,VL05,AKWTH05,GRGRH07}. This
interest is due to the technological advances in nanofabrication.
In addition, one of the most important recent development in
semiconductor has been the achievement of structures in which the
electronic behavior is essentially 2D. This means that the
charge carriers are confined in a potential such that their motion
in one direction is restricted and thus is quantized, leaving only
a two-dimensional momentum. In particular, there has been
considerable amount of work in recent years on semiconductor
confined structures, which finds applications in electronic and
optoelectronic devices. The application of a magnetic field
perpendicular to the heterostructure plane quantizes the energy
levels in the plane, drastically affecting the density of states
giving rise to the famous quantum Hall effect~\cite{Hall}. The
latter remains as the most interesting phenomenon observed in
physics because of its link to different theories and subjects.

Graphene~\cite{graph,graph2}, two-dimensional crystalline
materials, has became one of the most important subjects in
condensed matter research in the last few years. This new material
has a number of unique properties, which makes it one of the most
promising materials for future nanoelectronics~\cite{GN07}. One of
them is the band structure, which is gapless and exhibits a linear
dispersion relation at two inequivalent points K and K' in the
vicinity of the Fermi energy. Moreover, its low energy of
electrons are governed by a (2+1) dimensional Dirac equation.
Those electrons behave as massless chiral fermions,
{i.e.} relativistic electrons.
Consequently, the electrons cannot be localized by any confinement
potential, which is related to the fact that electrons in graphene
can have both positive and negative energies, {i.e.} Klein
tunneling effect~\cite{KNG06}. Graphene quantum dot (QD)~\cite{C99,SE07,MP08}
have sparked intense research activities related to quantum
information storage and processing using the spin information of
confined electrons. Various methods were used to make QD one of
the most widely-used techniques using electrostatic
gates~\cite{SE07}.

%Motivated by different investigations on the Dirac fermions in
%(2+1) dimensions~\cite{BJZ14,JAB09}, we use our developed
%formalism for the 2D Dirac equation~\cite{JAB09} to study the
%energy spectrum of {gate-tunable} graphene QD. Note
%that the QD is defined by gates introducing an electrostatic
%confining potential. We have obtained the solution of the Dirac
%equation in (2+1) dimensions in the presence of a constant
%magnetic field normal to the plane together with a two-dimensional
%Dirac-oscillator potential coupling.
On the other hand, the Dirac oscillator was
proposed by Moshinsky and Szczepaniak~\cite{MS89} in 1989 and is
considered as the relativistic version of the harmonic oscillator.
The Dirac oscillator has been studied
extensively~\cite{BMR90,RA99,N06,BML08,GKZ10,DPR13} because of
their probable applications in many branches of physics.
Additionally, the Dirac oscillator has been used in
optics~\cite{L10}, Jaynes Cummings model~\cite{STS10} and
grephene~\cite{QSxiv}. It is only recently that the first
experimental microwave realization of the one-dimensional Dirac
oscillator was developed~\cite{FSB13}.

We combine different approaches to achieve our goal. Indeed,
based on~\cite{JAB09,BJZ14} we set
%In this paper, we start by giving
the Hamiltonian system of our problem where a similarity
transformation is used to simplify the process for obtaining
the solutions. Later on, we define the QD by gates introducing an electrostatic
confining potential.
%We use the exact relationship between spinor
%components to obtain a second-order differential equation for one
%of the two-spinor components. Then
We find the bound state
solution of {gate-tunable} graphene QD in the
presence of a constant magnetic field $B$ and Dirac oscillator of frequency $\om$ as well as
a mass term that might be introduced by the
underlying substrate~\cite{GKB07,ZBF07}.
The eigenspinors are obtained in terms of the confluent hypergeometric functions showing
dependence of $B$ and the oscillator
coupling $\om$. %This allows to distinguish three different regimes according
%to the nature of

Subsequently, we analyze
the impact of the external field $B$ on the solutions of the energy spectrum of the QD
by extracting interesting properties. %different influence of an
%external magnetic field, perpendicular to the graphene layer, on
%the energy spectrum of the QD is also analyzed.
More precisely, we
consider three different cases corresponding to the relative
strength of $B$ with respect to $\om$. %the external magnetic field to the oscillator
%coupling.
In doing so, we start by defining
 an effective magnetic field, that
produces the effective quantized Landau levels, and focus on
%This suggests defining an effective magnetic field that
%produces the effective quantized Landau levels and
%. We focus on the
%effective magnetic field
its dependence of the bound states in circularly
symmetric QD. We show how to control the valley degeneracy by manipulating  the
effective magnetic field. This can help to form the valley
filters, valves~\cite{RTB07}, or qubits~\cite{RTRBB07}, and spin
qubits~\cite{TBL07} in graphene.

The paper is organized as follows. In section 2, we set our problem by reviewing
some mathematical tools need to deal with our issues.
To investigate the basic feature of the gate-tunable graphene QD, we set the appropriate confining
potential and
give the corresponding solutions of the energy spectrum in section 3.
Using the matching condition at the boundary, we obtain the condition that
governs the existence of the bound state. This will serve to study different liming cases related to the strength
of the magnetic field. We conclude our results in the final section.

%This will help to study the bound state appearing in our system and

%%%%%%%%%%%%%%%%%%%%%%%%%%%%%%%%%%%%%%%%%%%%%%%%%%
\section{Theoretical model}
%%%%%%%%%%%%%%%%%%%%%%%%%%%%%%%%%%%%%%%%%%%%%%%%%%%%%%

In order to deal with our task we establish an appropriate Dirac equation
describing our system. To go deeply in our study for the graphene QD, we
introduce a mass term to open a gap.

%%%%%%%%%%%%%%%%%%%%%%%%%%%%%%%%%%%%%%%
\subsection{Dirac equation}
%%%%%%%%%%%%%%%%%%%%%%%%%%%%%%%%%%%%%%

To start let us set some mathematical background
related to Dirac formalism needed to deal with our task. Indeed,
a particle of mass $m$ in the
presence of a constant perpendicular magnetic field can be described by
consider the Dirac equation in $(2+1)$ dimensions
\begin{equation}\label{eq1}
\left[ {{{i}}{\gamma ^\mu }({\partial _\mu } + {{i}}{A_\mu }) -
%\tau
m} \right]\psi  = 0, \qquad \mu =0,1,2
\end{equation}
where  the electromagnetic potential ${A_\mu } =({A_0},\vec A)$ and the space-time gradient
$\partial _\mu=\left(\frac{\partial}{\partial t},\vec \nabla
\right)$. Here we have the representation ${\gamma ^0} = {\sigma _3}$ and $
\vec \gamma  = {{i}}\,\vec \sigma$
with
the $2\times 2$ hermitian Pauli spin matrices $\left\{ {{\sigma _i}} \right\}_{i= 1}^3$.
%$\tau=\pm$ differentiates the two valleys $K$ and $K'$.
The Dirac matrices ${\gamma ^\mu }$ satisfy the algebra
%= \left( {{\gamma ^0},\vec \gamma } \right)$ are
%three unimodular square matrices satisfying the anti-commutation
%relation
\begin{equation}
\left\{ {{\gamma ^\mu },{\gamma ^\nu }} \right\} = {\gamma ^\mu
}{\gamma ^\nu }
 + {\gamma ^\nu }{\gamma ^\mu } = 2{{\cal G}^{\mu
\nu }}, \qquad \mu, \nu =0,1,2
\end{equation}
 with the metric ${\cal G}=\mbox{diag} ( + \,\, -
\,\,- )$. % and $\mu , \nu =0,1,2$.
 %is the metric of
%Minkowski space-time, which is equal to $\mbox{diag} ( + \,\, -
%\,\,- )$.
% A minimal irreducible matrix representation that
%satisfies this relation is taken as ${\gamma ^0} = {\sigma _3}$, $
%\vec \gamma  = {{i}}\,\vec \sigma$
% where $\left\{ {{\sigma _i}} \right\}_{i= 1}^3$
%are the $2\times 2$ hermitian Pauli spin matrices
%:
%\begin{equation}
%{\sigma _1} = \left( {\begin{array}{*{20}{c}}
%0 & 1  \\
%1 & 0  \\
%\end{array}} \right),
%\qquad {\sigma _2} = \left( {\begin{array}{*{20}{c}}
%0 & -i  \\
%i & 0  \\
%\end{array}} \right),\qquad
%{\sigma _3} = \left( {\begin{array}{*{20}{c}}
%1 & 0  \\
%0 & -1  \\
%\end{array}} \right).
%\end{equation}

In our study of system made of graphene, we need to consider massless Dirac fermions. To this end,
we multiply  \eqref{eq1} by $\si_3$ to open a gap, such as
%then we need
%For later purpose it is convenient to write \eqref{eq1} in the form
\begin{equation}
{{i}}{{\partial  \over {\partial t}}}\psi
 = \left( { - {{i}}\,\vec \alpha  \cdot \vec \nabla  + \vec \alpha
\cdot \vec A + {A_0} + m\beta } \right)\psi \label{eq3}
\end{equation}
where we have set
%$\vec \alpha $ and $\beta$ are the hermitian matrices:
$\vec
\alpha  = {{i}}\,{\sigma _3}\vec \sigma$ and $ \beta  = {\sigma _3}$.
%We will see below that the symmetry of the problem is preserved even
%if.
%We introduce an oscillator coupling in Equation~\eqref{eq1}
For time independent
potentials, the two-component spinor wavefunction is separable
%$\psi(t,r,\theta )$ is written as
%\begin{equation}
$\psi (t,r,\theta ) = {e^{- {{i}}E t}}\psi (r,\theta ).$
%\end{equation}
For regular solutions of \eqref{eq3}, square integrability
%(with respect to the measure ${d^2}\vec r = r\,dr\,d\theta $)
and the boundary conditions require that $\psi (r,\theta )$
satisfies
\begin{equation}
{\left. {\sqrt r \,\psi (r,\theta )} \right|_{\scriptstyle r = 0
\hfill \atop \scriptstyle r \to \infty  \hfill}} = 0, \qquad \psi
(\theta+2\pi) = \psi (\theta). \label{eq6}
\end{equation}
To simplify the construction of the solution, we look for a local
$2\times2$ similarity transformation $\Lambda (r,\theta )$ that
maps the cylindrical projection of the Pauli matrices $(\vec
\sigma  \cdot \hat r$, $\vec \sigma  \cdot \hat \theta )$ into
their canonical Cartesian representation $({\sigma _1}, {\sigma
_2})$, respectively. That is
\begin{equation}
\Lambda\, \vec \sigma  \cdot \hat r\,{\Lambda ^{ - 1}} = {\sigma
_1}, \qquad \Lambda \,\vec \sigma  \cdot \hat \theta\,{\Lambda ^{
- 1}} = {\sigma _2}. \label{eq7}
\end{equation}
We note that any other choice for the pair of Pauli matrices can
be obtained from the present one through a unitary transformation,
hence leaving the physics of the problem unaltered. A $2\times 2$
matrix that is defined by~\cite{JAB09,BJZ14}
%satisfies this requirement is
\begin{equation}
\Lambda (r,\theta ) =\frac{1}{\sqrt{r}}
%\lambda (r,\theta )
\,{e^{{{}{{{i}}
\over 2}}{\sigma _3}\theta }}. %,\qquad
\label{eq8}
\end{equation}

We are interested to
the Dirac oscillator for its probable application in many
branches of physics as we noticed before \cite{L10,QSxiv,FSB13}.
%such as in optics~\cite{L10},
%graphene~\cite{QSxiv} etc. Recently it has also been realized
%experimentally in microwaves~\cite{FSB13}.
Motivated by these investigations, we % motivated us to
consider such oscillator in another context and emphasize its influence on
a system based on the QD. %from another point of view that is related to the QD.
To achieve this goal, we
introduce an additional coupling as the 2D Dirac-oscillator
potential~\cite{MS89,MS96}, that keeps symmetry of the system.
This coupling is introduced by the substitution $ \vec \nabla
 \to \vec \nabla  + \lambda\omega \vec r\beta$
where $\omega$ is the oscillator frequency and $\lambda$ is a constant parameter.
To simplify the forthcoming analysis, we require that the condition $\lambda=m$ should be fulfilled.
Now from the above consideration, we
%Finally, we
obtain the $(2+1)$-dimensional Dirac equation for a
charged spinor in static electromagnetic potential
 \beq\lb{eq777}
 \left( {H
- E } \right)\chi_\pm  = 0 \eeq
 where the Hamiltonian is given by
\begin{equation}
H=\left( {\begin{array}{*{20}{c}} { 0 } & {{\partial _r}
+ {{i}}{A_r} - \lam\omega r - {{{{i}} \over r}}{\partial _\theta } +
{A_\theta }}  \\
{ - {\partial _r} - {{i}}{A_r} - \lam\omega r - {{}{{{i}} \over
r}}{\partial _\theta } + {A_\theta }} & { 0 }
%\\
%\end{array}} \right)\left( {\begin{array}{*{20}{c}}
%{\mathop {{\chi _ + }(r,\theta )}\limits_{} }  \\
%{{\chi _ - }(r,\theta )}  \\
\end{array}} \right)
+\lam\si_3 +A_0 \mathbb{I}
\label{eq13}
%(2.12)
\end{equation}
%where
and ${\chi_\pm }$ are the components of the transformed
wavefunction $\left| \chi  \right\rangle  = \Lambda \left| \psi
\right\rangle $, with $\Lambda$ is given in \eqref{eq8}. It is clearly seen that
the second term is gap and third one can be regarded as an external potential. In the forthcoming analysis, we will %see how {to}
fix different potential in order to deal with some basic features
some properties of
the gate-tunable graphene QD.

%%%%%%%%%%%%%%%%%%%%%%%%%%%%%%%%%%%%%%%%%%%%%%
\subsection{Mass term}
%%%%%%%%%%%%%%%%%%%%%%%%%%%%%%%%%%%%%%%%%%%%%%%

To study the impact of the external field $B$ and oscillating frequency
$\om$
we consider a system made of graphene described by the Hamiltonian %whose corresponding
%Hamiltonian
%The Hamiltonian that describe our system
%is given in
\eqref{eq13} except we replace $\lam \sigma_3$ by $\tau
\lam\sigma_3$ with $\tau=\pm1$ differentiates the two valleys $K$ and
$K'$.
%{So, the Hamiltonian defined in equation
%\eqref{eq13} will depend on $\tau$.
Thus, we call the wave
function $\chi_{\pm}^{\tau}$ spinor. To go further, let us set
some quantities such
%introduce
a constant magnetic field of strength $B$
applied perpendicular to the
%at right angles
($r, \theta $)-plane, which is $\vec B = B\,\hat z$. We choose the
gauge
%Therefore, the electromagnetic potential has the time and space
%components:
%\begin{equation}
%{A_0} = 0, \qquad
$\vec A(r,\theta ) = \frac{B}{2}\left(0,r\right)$
% \,\hat \theta.
%\end{equation}
and assume a circular symmetry in the confinement potential ${A_0}
= U(r)$.

Consequently, \eqref{eq777}
%(2.12)
becomes completely separable in radial and angular parts. Then, we can write the spinor
wavefunction as
\begin{equation}
{{\chi_{\pm}^{\tau} }}(r,\theta ) =
{{\phi_{\pm}^{\tau}} }(r)\,\varphi (\theta )
\end{equation}
such that
the angular component satisfies the eigenvalue equation $ - {{i}}{{}{{d\varphi }
\over {d\theta }}} = \xi \,\varphi $ where {$\xi{}$} is a real
separation constant giving the function
\begin{equation}
\varphi (\theta ) = {{1 \over {\sqrt {2\pi } }}}{e^{{{i}}\xi
\theta }}.
\end{equation}
On the other hand, the boundary condition $\psi (\theta  + 2\pi )
= \psi (\theta)$ requires the following condition
%\begin{equation}
\beq
{e^{{{i}}\,2\pi \xi }}{e^{ - {{i}}{\sigma _3}\pi }} =  + 1
\eeq
which, in turn, demand that $ {e^{{{i}}\,2\pi \xi }} =  - 1 $
giving the quantum number
\beq
\xi  = {{1 \over 2}}\kappa, \qquad
\kappa=\pm 1, \pm 3, \pm 5 \cdots.
\eeq
%%%%%%%%%%%%%%%%%%%%%%%%%%%%%%%%%%%%%%%%%%%%%%%%%%%%%%
%\subsection{Bound-state solutions}
%%%%%%%%%%%%%%%%%%%%%%%%%%%%%%%%%%%%%%%%%%%%%%%%%%%%%%
%Consequently,
While, the Dirac equation for the two-component radial spinor is reduced
to
\begin{equation}\left( {\begin{array}{*{20}{c}}
{\tau \lam +U - E } & {{{d \over {dr}}} +
{{\xi  \over {r}}} + Gr}  \\
{ - {{d \over {dr}}} + {{\xi  \over {r}}} +
Gr} & { - \tau \lam +U - E }  \\
\end{array}} \right)\left( {\begin{array}{*{20}{c}}
{\mathop {{{\phi_{+}^{\tau}} }(r)}\limits_{} }  \\
{{{\phi_{-}^{\tau}} }(r)}  \\
\end{array}} \right) = 0\label{eq15}
%(2.14)
\end{equation}
where the physical constant %$G$ is given by
$ G = \lambda\left(\omega_c - \om\right)$ and  the cyclotron
frequency $\om_c={{B \over 2\lambda}}$. Thus, the presence of the 2D
Dirac-oscillator coupling did, in fact, maintain the symmetry of
the problem. It is interesting to note that the effect of the
oscillator interaction is to produce an effective magnetic field.
This effective magnetic field is assumed to originate in an
effective form from the motion of the charge carriers relative to
the planar hexagonal arrangement of carbon atoms. Moreover, its
introduction is equivalent to change the magnetic field as
\beq
B
\lga B - 2\lambda\omega.
\eeq
 As a result of
the wave equation \eqref{eq15}, the two spinor components satisfy
the relation
\begin{equation}
{{\phi_{\mp}^{\tau}} }(r) = \frac{1}{{E-U  \pm \tau
\lam}}\left[ { \mp \frac{d}{{dr}} + \frac{\xi} {r} + Gr}
\right]{{\phi_{\pm}^{\tau}} }(r) \label{eq16}.
%(2.15)
\end{equation}
%where $E  \ne  \pm \tau m -U$.
%Therefore, the solution of the
%problem with the top/bottom sign corresponds to the
%positive/negative energy solution.
Using  \eqref{eq16}
%(2.15)
to eliminate one component in terms of the other in (\ref{eq15}), which
results in the following Schr\"{o}dinger-like differential
equation for each spinor component
\begin{equation}
\left\{ { - \frac{{{d^2}}}{{d{r^2}}} + \frac{{{{}{\xi}}\left(
{{{}{\xi}} \mp 1} \right)}}{{{r^2}}} + {G^2}{r^2} + \left[ {{ \lam^2}
- {(E-U) ^2} + G\left( {2\xi  \pm 1} \right)} \right]}
\right\}{{\phi_{\pm}^{\tau}} }(r) = 0. \label{eq17}
\end{equation}
%It may be noted {that}
%equation
%\eqref{eq17} does note depend on the
%valley index $\tau$.
It is clearly seen that this equation can completely be solved by choosing an approriate form of
 the potential $U(r)$. This will be done
by specifying the nature of our system, which be will be the graphene QD one. However, for a constant potential
and on the light of the above choice, we show that he solutions of
\eqref{eq17}  take the form %will be of the form
\begin{equation}
\phi_{\sigma}^{\tau}(r) \sim e^{-\left| G
\right|r^2/2} r^{\left(\left| \xi-\frac{\sigma}{2}
\right|+1/2\right)}\Gamma(r)
\end{equation}
 and  the function $\Gamma(r)$ is given by the following combination
 \beq
 \Gamma(r)=\alpha_\sigma U\left(n_\sigma, \left|
\xi-\frac{\sigma}{2} \right|+1,\left| G \right|r^2\right)+\beta_\sigma
M\left(n_\sigma, \left| \xi-\frac{\sigma}{2} \right|+1,\left| G
\right|r^2\right)
\eeq
with $U$ and $M$ are the confluent hypergeometric
functions. The parameters $\alpha_\sigma$ and $\be_\sigma$ are normalization constants,
which can be fixed by the boundary conditions. We show that
the corresponding eigenvalues $\epsilon=E-U$ read as
\beq
 n_\sigma= \frac{\lam^2-\epsilon^2}{4\left| G \right|}
+\frac{1}{2}\left(1+\left| \xi-\frac{\sigma}{2}
\right|+s\left(\xi+\frac{\sigma}{2}\right)\right)
 \eeq
 where $n_\sigma$ is an integer value and $s=\mbox{sgn} G$. We notice that teh above results are in agreement
 with those reported in \cite{JAB09} for $\si=1$.
 % The quantum number $(n=0,\, 1,\, 2,
%\,\cdots)$ is given by

%%%%%%%%%%%%%%%%%%%%%%%%%%%%%%%%%%%%%%%%%%%%%%
\section{Graphene quantum dot}
%%%%%%%%%%%%%%%%%%%%%%%%%%%%%%%%%%%%%%%%%%%%%%%

We show that how the above results can be used to study some physical properties of
a system made of the graphene QD.
This latter is defined by gates
introducing an electrostatic confining potential
as shown in Figure \ref{PGp}. In what follow
we study the corresponding solutions in the presence of a
constant mass term $\lam$, that will account for a gap $2\lam$ in the
energy spectrum. In fact, we will consider three limiting cases
those related to the strength of the external magnetic field with respect to Dirac
oscillating frequency.

%%%%%%%%%%%%%%%%%%%%%%%%%%%%%%%%%%%%%
\subsection{Potential configuration}
%%%%%%%%%%%%%%%%%%%%%%%%%%%%%%%%%%%%%%%%%

\begin{figure}[h!]
\centering
\includegraphics[width=7cm, height=5cm]{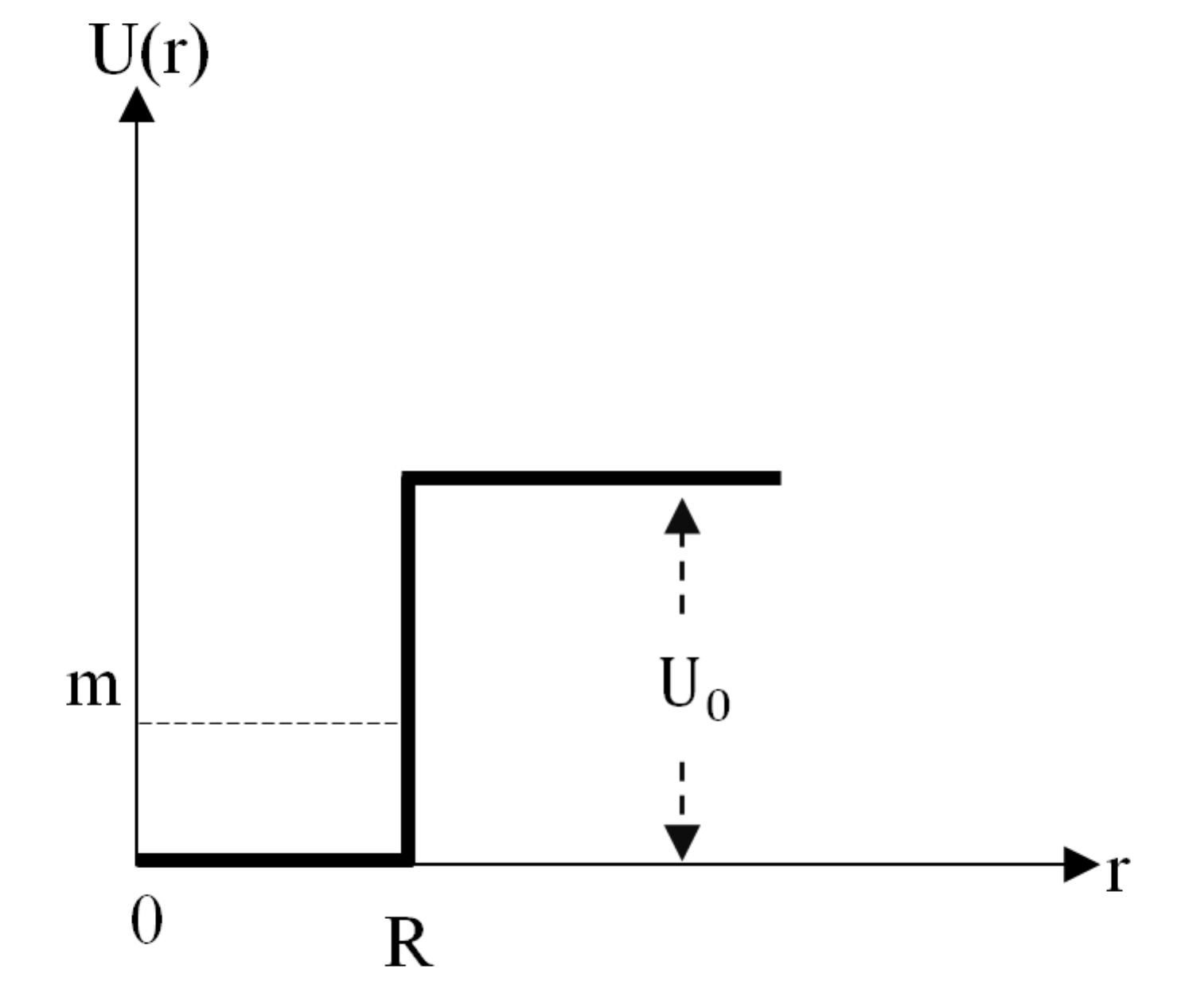}
\caption{ {\sf {Potential landscape of a QD with radius $R$.}}}\label{PGp}
\end{figure}

We are interested by studying the impact of the Dirac oscillator coupling together with
an external magnetic field on the graphene QD. In doing so, we look first for
%We need to find
the bound states solution for the QD governed by the second order
differential equation \eqref{eq17}.
To this end, we take a radially symmetric model with the following
potential
%For this we use the following confinement potential
\begin{equation} \label{eq:Uform}
U(r) =  \left
\{ \begin{array}{cc} U_0, &  \; r>R \;
\\ 0, &  \; r<R \;
\end{array} \right.
\end{equation}
and its profile, %of the potential,
which consists of two
regions, is schematically shown in Figure \ref{PGp}.
The radially
symmetric choice makes the analytic determination of the bound state solution
possible.\\

According to Figure \ref{PGp}, we  distinguish between two regions in order
to explicitly determine the solutions of the energy spectrum. Consequently,
the general solutions to the radial equation %Dirac equation
\eqref{eq17}, which are
regular at the origin and  decay exponentially as
$r\rightarrow \infty$, are given in terms of $ M\left(n_\sigma, \left|
\xi-\frac{\sigma}{2} \right|+1,\left| G \right|r^2\right)$ inside the QD
and $ U\left(n_\sigma, \left| \xi-\frac{\sigma}{2} \right|+1,\left| G
\right|r^2\right)$ outside the QD. These are
\beq \label{ansatz1}
{{\phi_{\sigma}^{\tau}}}(r)=2^{\left(1+\left|
\xi-\sigma/2 \right|\right)/2} e^{-\left| G  \right|r^2/2}
r^{\left(\left| \xi-\frac{\sigma}{2} \right|+1/2\right)}  \left\{
\begin{array}{cc} \alpha_\sigma U\left({n_{\sigma}^{>}}, \left|
\xi-\frac{\sigma}{2} \right|+1,\left| G \right|r^2\right), &  r>R
\\  \beta_\sigma M\left({n_{\sigma}^{<}},
\left| \xi-\frac{\sigma}{2} \right|+1,\left| G  \right|r^2\right), & r<R
\end{array} \right. \eeq
%\end{multline}
where $\sigma=\pm 1 $ the upper sign corresponding to the A
sublattice and the lower sign to the B one. We introduced
the label $<,>$ to separate between {%Here we have introduced
 eignenergies of the two regions such that
 $\epsilon_< \equiv E$ and $\epsilon_> \equiv E- U_0$.
Note that, the absolute value of $G$ is introduced in order to consider % as we considered
the case of
positive and negative effective magnetic field.}
%, respectively.
In addition, we note that the solution
${\phi_{\sigma}^{\tau}}(r)$ depend on the valley
index $\tau$ through \eqref{eq15}. {Such that
$\phi_{+}^{+}(r)$ and $\phi_{-}^{+}(r)$ are envelope wave
functions of A and B sublattices for the valley $K$, and
$\phi_{+}^{-}(r)$, $\phi_{-}^{-}(r)$ for the valley $K'$.}

At this stage,  it is
natural to ask about the eigenvalues associated to the bound states of
our system.  To answer
this inquiry we use the matching conditions of the spinors at the
interface $r=R$ to end up with the  condition
for the existence of bound {states}
\beq
\label{cara.eq} \frac{\alpha_+}{\alpha_-}\
\frac{U \left(n_{+}^{>},|\xi-\frac{1}{2}|+1,|G|R^2\right)}{U\left(n_{-}^{>},|\xi+\frac{1}{2}|+1,|G|R^2\right)}=\frac{\beta_+}{\beta_-}\
\frac{M \left(n_{+}^{<},|\xi-\frac{1}{2}|+1,|G|R^2\right)}{M\left(n_{-}^{<},|\xi+\frac{1}{2}|+1,|G|R^2\right)}
\eeq
where the ratios $\alpha_{+}/\alpha_-$ and $\beta_+/\beta_-$
can be explicitly determined by making use of the coupled first-order differential equation
\eqref{eq16}.
%We further introduce the parameters
%\beq
% n_\sigma^{<,>}= \frac{m^2-\epsilon_{<,>}^2}{4\left| G \right|}
%+\frac{1}{2}\left(1+\left| \xi-\frac{\sigma}{2}
%\right|+s\left(\xi+\frac{\sigma}{2}\right)\right)
% \eeq
%with $ s= \mbox{sgn}(G)$
%and we define the corresponding energies as $\epsilon_< \equiv E$
%and $\epsilon_> \equiv E- U_0$.
The sign of $G$ depends on whether
the oscillator frequency $\omega$ is larger or smaller than the
cyclotron frequency $\omega_c$.

%In this context,
It is worthwhile
investigating the basic features of %some limits
of the above results and
underlying their interesting properties.
In this context, we
would like to show how the valley degeneracy can be lifted by the
presence of a constant magnetic field normal to the plane of the system together
with a 2D Dirac-oscillator. %  coupling.
This
is of particular importance to form the valley filters,
valves~\cite{RTB07}, or qubits~\cite{RTRBB07} and spin
qubits~\cite{TBL07} in graphene.
To this end, we consider
three different cases corresponding to the relative strength of
the magnetic field ($\omega_c$) with respect to the oscillator coupling
($\omega$).

%%%%%%%%%%%%%%%%%%%%%%%%%%%%%%%%%%%%%%%%%%%%%%%%%%%%%%%%%%%%%%%
\subsection{Oscillating with same frequency}
%%%%%%%%%%%%%%%%%%%%%%%%%%%%%%%%%%%%%%%%%%%%%%%%%%%%%%%%%%%%%%%%%

We study the first case were the oscillator frequency is
tuned to resonate with the cyclotron frequency, i.e. $w\approx
w_c$, which it is equivalent to consider $G=\lambda \Delta'$ with $\Delta'=\omega_c
-\omega$. By requiring %the following constraint
${|\Delta'| \ll \lambda}$ %. In this
%limit (for small $G$),
we show that the hypergeometric functions reduce to
bessel functions~\cite{Abramo}
\beqar
&& U\left(n^{>},j,z\right)=\frac{2}{\Gamma\left(1-j+n^{>}\right)}(z
n^{>})^{(1-j)/2} K_{j-1}\left(2 \sqrt{z n^{>}}\right) \label{abramo2}  \\
&&  M\left(n^{<},j,z\right)=\Gamma(j) \left(-z n^{<}\right)^{(1-j)/2}
J_{j-1}\left(2 \sqrt{-z n^{<}}\right) \label{abramo1}
\eeqar
%For $G=0$,
%this allow to
and  the characteristic equation \eqref{cara.eq} takes the form according
to
$\xi
> 0$
%is giving by
 \beq
\label{eq1.G0}\frac{\sqrt{ \lam^2-(E-U)^2}}{\sqrt{E^2- \lam^2}}
\frac{K_{\xi-\frac{1}{2}}\left(\sqrt{
\lam^2-(E-U)^2}R\right)}{K_{\xi+\frac{1}{2}}\left( \sqrt{
\lam^2-(E-U)^2}R\right)}=-\frac{E-U-\tau \lam }{E-\tau \lam}
\frac{J_{\xi-\frac{1}{2}}\left(\sqrt{
E^2-\lam^2}R\right)}{J_{\xi+\frac{1}{2}}\left(\sqrt{ E^2-
\lam^2}R\right)}
\eeq
or $\xi < 0$ %, \textcolor{red}{we obtain}
\beq \label{eq2.G0}
\frac{\sqrt{\lam^2-(E-U)^2}}{\sqrt{E^2-\lam^2}}
\frac{K_{-\xi-\frac{1}{2}}\left(
\sqrt{\lam^2-(E-U)^2}R\right)}{K_{-\xi+\frac{1}{2}}\left(\sqrt{
\lam^2-(E-U)^2}R\right)}=-\frac{E-U+\tau \lam }{E+\tau \lam}
\frac{J_{-\xi-\frac{1}{2}}\left(\sqrt{
E^2-\lam^2}R\right)}{J_{-\xi+\frac{1}{2}}\left(\sqrt{
E^2-\lam^2}R\right)}.
\eeq
We notice that there is a mapping between both of the two last equations, which is assured by
the change
%\eqref{eq2.G0} can be obtained from
%equation
%\eqref{eq1.G0} by using the following
%map
$\xi\rightarrow -\xi$ and $\tau \rightarrow -\tau$. In general, these two equations
%both characteristic equation \eqref{eq1.G0} and \eqref{eq2.G0}
can
not be solved in the closed form.

\begin{figure}[h!]
\centering
\includegraphics[width=7cm, height=7cm]{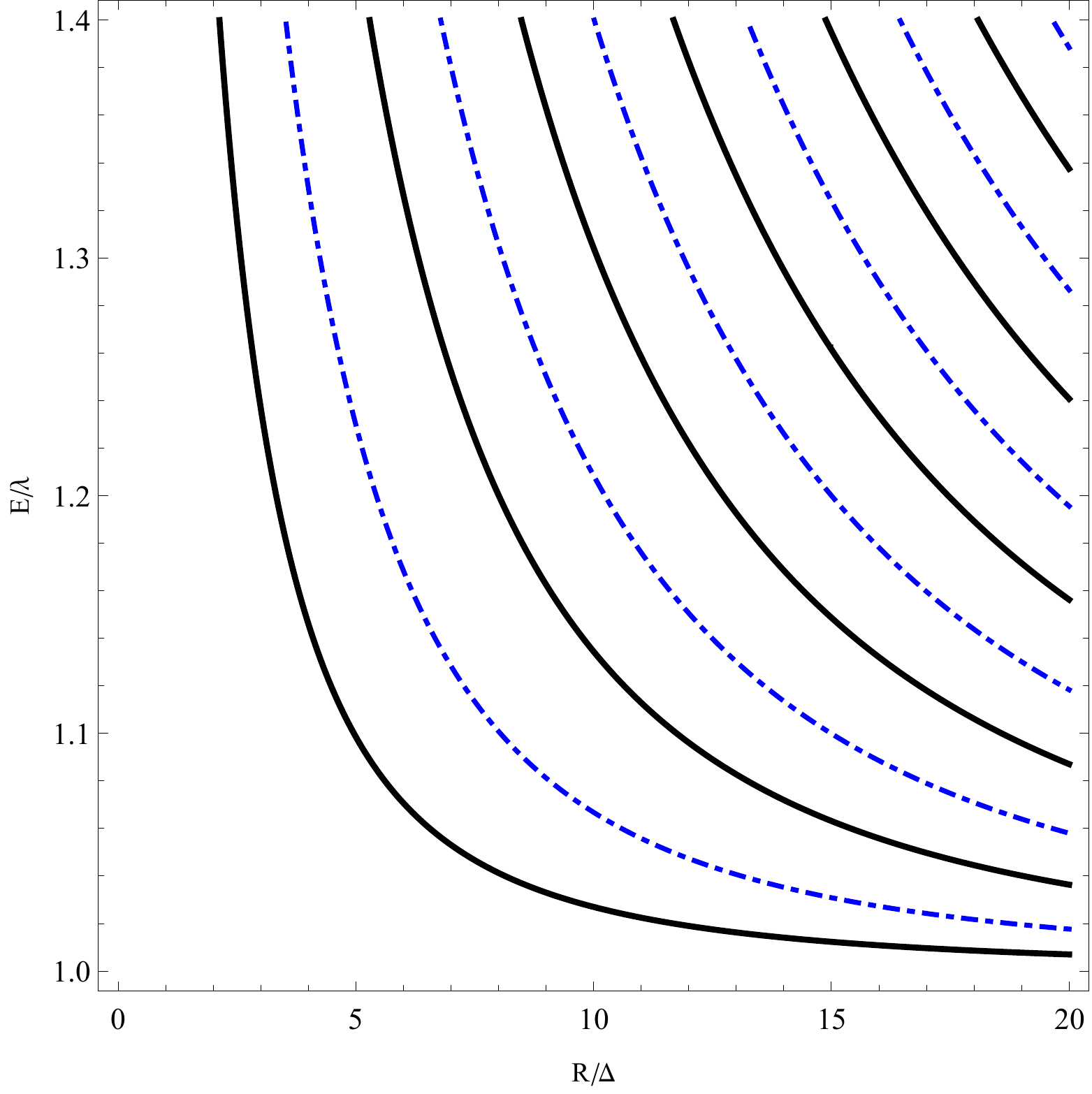}
\caption{ {\sf {Energy levels of a circular quantum dot as a
function of the dot radius $R$ for $\xi=\frac{1}{2}$ {and}
%with
$U=1.5\ \lam$, {with} %where
$\Delta= \hbar v_F/\lam$. The energy levels corresponding to the $K$
and $K'$ valleys are shown by the solid curve ($\tau=1$) and
the blue dot dashed curve ($\tau=-1$) {,
respectively}.}}}\label{G=0}
\end{figure}

Results for the energy levels as a function of the dot radius are
shown in Figure~\ref{G=0} for $\xi=\frac{1}{2}$
{and} $U=1.5\ \lam$. The black solid and blue dot
dashed curves correspond to the $K$ and $K'$ valleys,  respectively.
Note that the {case} %for
$\xi=-\frac{1}{2}$ for the same valley is also presented in both
curves in Figure~\ref{G=0}, which can be discussed by taking into account of the symmetry
$E(\xi,\tau)= E(-\xi,-\tau)$ exhibited by our system. It may be noted that the solution
for the two curves are different for the same valley. This can be
explained by the fact that for the same valley we have $E(\xi)\neq
E(-\xi)$. In fact, we can easily see {from}
%equation
 \eqref{eq1.G0} and \eqref{eq2.G0} that the energy verify
$E(\xi,\tau)= E(-\xi,-\tau)$ and $E(\xi,\tau)\neq E(-\xi,\tau)$,
which helps us to study the valley degeneracy by a Dirac
oscillator. However, the results show that the valley degeneracy
is not broken, in this case, if we include both sign of $\xi$. Our
results are in agreement with previous calculations of bound-state
energies for zero magnetic field~\cite{Nilsson}.
In summary, it is clearly seen that
%{We note} that
when the oscillator frequency is
tuned to resonate with the cyclotron frequency, the Dirac
oscillator annihilates the effect of the magnetic field on our system and we find ourselves in
the case of zero effective magnetic field. At zero effective
magnetic field, the degeneracy of the levels in the two valleys is
clearly displayed.

%%%%%%%%%%%%%%%%%%%%%%%%%%%%%%%%%%%%%%%%%%%%%%%%%%
\subsection{Strong magnetic field}
%%%%%%%%%%%%%%%%%%%%%%%%%%%%%%%%%%%%%%%%%%%%%%%%%%%

Now, we study the case when the cyclotron frequency is much larger
than the oscillator frequency, i.e. $w_c\gg
w$, then $s=1$ and $G\approx \lambda\om_c$. It may be
noted that the effect of the oscillator interaction is to produce
an effective magnetic field along the positive $z$-direction that
produces the effective quantized Landau levels.

In this case we arrive, after
applying the boundary conditions and finding both ratios, at the
following characteristic equation for $\xi > 0$
\beq \label{eqS.car1} \frac{E-U+\tau \lam}{4G}\
\frac{U(n_1^>+\xi+\frac{1}{2},\xi+\frac{1}{2},G
R^2)}{U(n_1^>+\xi+\frac{1}{2},\xi+\frac{3}{2},G R^2)}=\frac{\xi
+\frac{1}{2}}{E-\tau \lam} \
\frac{M(n_1^<+\xi+\frac{1}{2},\xi+\frac{1}{2},G
R^2)}{M(n_1^<+\xi+\frac{1}{2},\xi+\frac{3}{2},G R^2)}
\eeq
and for
$\xi < 0$
 %\textcolor{red}{\eqref{cara.eq} becomes}
\beq \label{eqS.car2} \frac{1}{E-U+\tau \lam} \
\frac{U(n_1^{>},-\xi+\frac{1}{2},G
R^2)}{U(n_1^{>}+1,-\xi+\frac{3}{2},GR^2)}=\frac{
\xi-\frac{1}{2}}{E+\tau \lam} \ \frac{M(n_1^{<},-\xi+\frac{1}{2},G
R^2)}{M(n_1^{<}+1,-\xi+\frac{3}{2},GR^2)}
\eeq
where the used quantum number is given by
\beq
n_1^{<,>}=\frac{\lambda^2-\epsilon_{<,>}^2}{4|G|}.
\eeq

\begin{figure}[H]
\centering
\includegraphics[width=7cm, height=7cm]{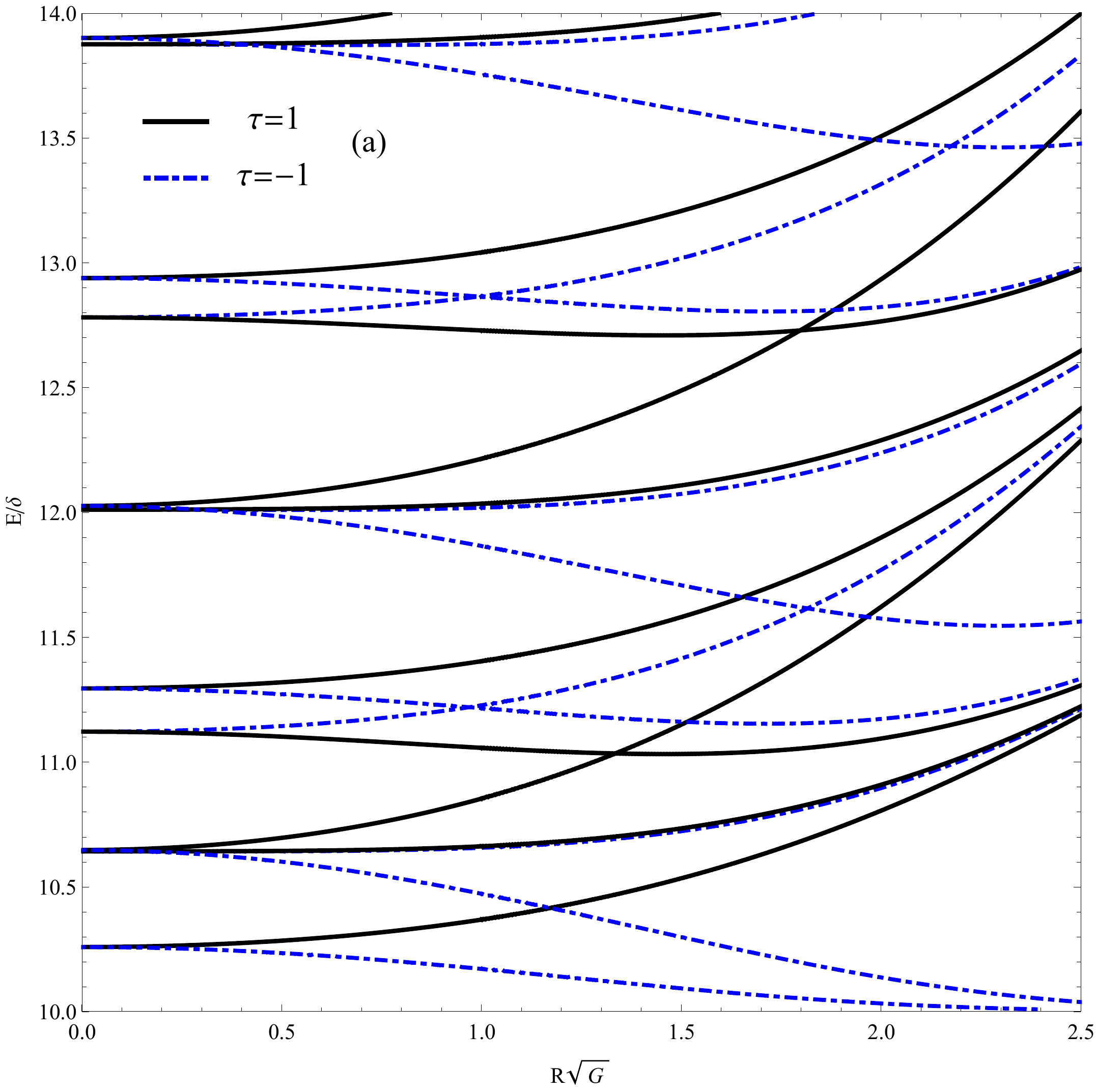} \ \ \ \ \ \ \ \ \ \ \
\includegraphics[width=7cm, height=7cm]{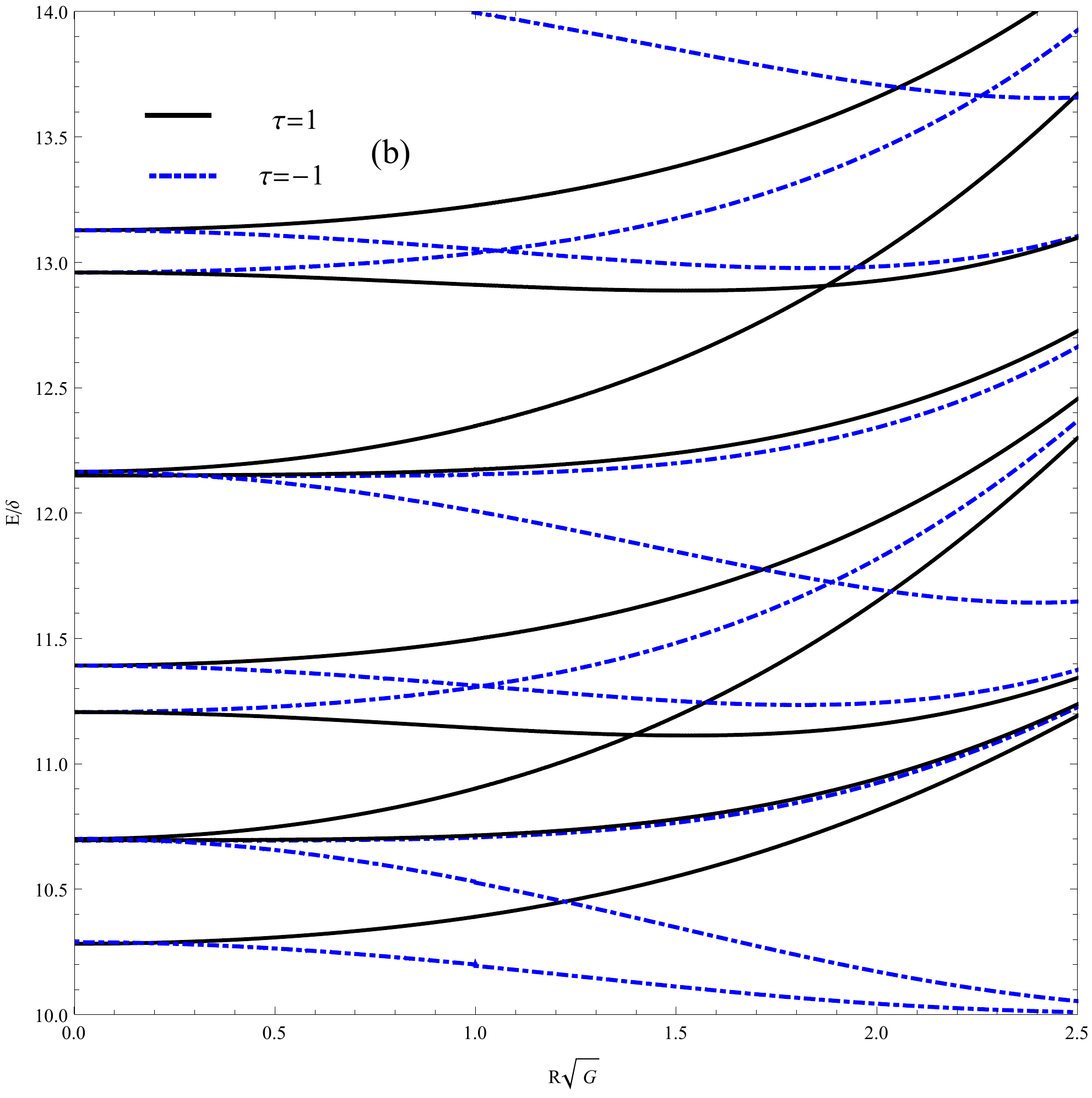}
\caption{ {\sf {Energy levels of a circular quantum dot as a
function of the dot radius $R$ with {$\lam=10\
\delta$, $U=10\ \delta$/ $U=20\ \delta$} ((a)/(b)) and we have
introduced the quantum dot level spacing $\delta=\hbar v_F /R$.
The energy levels corresponding to the $K$ and $K'$ valleys are
shown by the solid curve ($\tau=1$) and the dot dashed curve
($\tau=-1$){, respectively}. We show the low-lying
bound states for $\xi=\pm \frac{1}{2},\ \pm
\frac{3}{2}$.}}}\label{Gp}
\end{figure}

Figure~\ref{Gp} shows the energies of the bound states in a
graphene QD as function of the dot radius $R$ for small effective
magnetic field. In this Figure, we show the low-lying bound states
for $\xi=\pm \frac{1}{2},\ \pm \frac{3}{2}$ with $U=10
\delta$/$U=20 \delta$ (Figure \ref{Gp} (a)/(b)) and $\lam=10 \delta$.
We can clearly show that for zero effective magnetic field the
degeneracy of the levels in the two valleys is not broken. This is
in agreement with the results found in the first case when the
oscillator frequency is tuned to resonate with the cyclotron
frequency.
%We can clearly shown that
%when the oscillator frequency is tuned to resonate with the
%cyclotron frequency, that correspond to a zero effective magnetic
%field, the degeneracy of the levels in the two valleys is not
%broken.
However, by increasing the effective magnetic field, the valleys
degeneracy is broken. This allows us to conclude that the
oscillator interaction, which allow to produce an effective
magnetic field, can be used to control the orbital degeneracy. In
addition, we show that the number of the bound states depend on
the electrostatic potential. In fact, by increasing $U$ the number
of bound states decrease.\\

\begin{figure}[H]
\centering
\includegraphics[width=8cm, height=8cm]{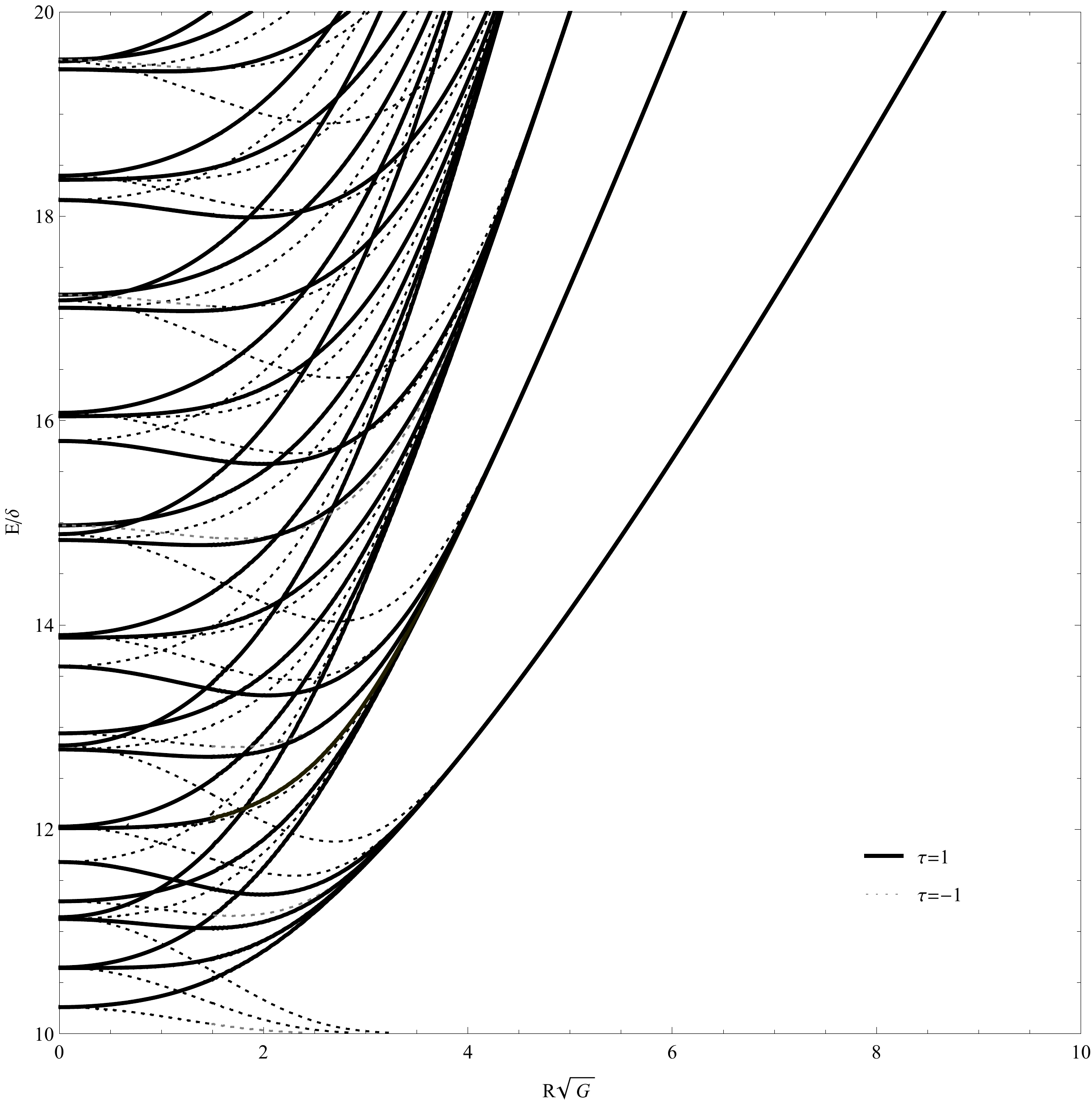}
\caption{ {\sf {Energy levels of a circular quantum dot as a
function of the dot radius $R$ with {$m=10\
\delta$}, $U=\lam$ and we have introduced the quantum dot level
spacing $\delta=\hbar v_F /R$. The energy levels corresponding to
the $K$ and $K'$ valleys are shown by the solid curve ($\tau=1$)
and the dot dashed curve ($\tau=-1$) respectively. We show the
bound states for large effective magnetic field with $\xi=\pm
\frac{1}{2},\ \pm \frac{3}{2},\ \pm \frac{5}{2}$.}}}\label{Gp2}
\end{figure}

In Figure~\ref{Gp2} we show the energies of the bound states in a
graphene QD as function of the dot radius $R$ for large effective
magnetic field for $\xi=\pm \frac{1}{2},\ \pm \frac{3}{2},\ \pm
\frac{5}{2}$. The solid and dot dashed curve correspond
respectively to the $K$ and $K'$ valleys. The results show that
for large effective magnetic field the energy levels converge to
the bulk Landau Levels. {In addition,
%it may be noted that
the states} that are degenerate for zero effective magnetic field
correspond to opposite values of the angular momentum $\xi$ in
different valleys. We note that our results are in agreement with
previous calculations of bound-state energies in presence of
magnetic field applied perpendicular to the graphene
plane~\cite{Nilsson}.

%\begin{itemize}
%    \item {\textbf{Weak-field case} }%($w_c\ll w$)}
%\end{itemize}

%%%%%%%%%%%%%%%%%%%%%%%%%%%%%%%%%%%%%%%%%%%%%%%%%%%%%%%%%%%
\subsection{Weak magnetic field}
%%%%%%%%%%%%%%%%%%%%%%%%%%%%%%%%%%%%%%%%%%%%%%%%%%%%%%%%%%%%%%%%

In the third case we suppose that the cyclotron frequency is much
smaller than the oscillator frequency, namely $\omega_c \ll
\omega$, with $G\approx -\lam\omega$, or $s=-1$. It may be noted that in
this situation the effect of the oscillator interaction is to produce
an effective magnetic field along the negative $z$-direction. In
fact, this can be done by interchanging the confinement frequency
$\omega$ with the cyclotron one $\omega_c$.

By using the above consideration, we obtain from~\eqref{cara.eq}
the characteristic equation for the allowed eigenenergies $E$ of the QD. Indeed, for
$\xi > 0$ we have
%For $\xi > 0$, we obtain the following characteristic equation for
 %under consideration
\beq \label{eqW.car1} \frac{1}{E-U-\tau \lam}\
\frac{U(n^{>}_1,\xi+\frac{1}{2},-GR^2)}{U(n^{>}_1
+1,\xi+\frac{3}{2},-GR^2)}=\frac{- \xi-\frac{1}{2}}{E-\tau \lam} \
\frac{M(n^{<}_1,\xi+\frac{1}{2},-GR^2)}{M(n^{<}_1+1,\xi+\frac{3}{2},-GR^2)}
\eeq
and for $\xi < 0$ reads as
%, the characteristic equation~\eqref{cara.eq} for
%the weak-field case can be written in the following form
\beq \label{eqW.car2} \frac{E-U-\tau \lam}{4 G}\ \frac{U(n^>_1-
\xi+\frac{1}{2},-\xi+\frac{1}{2},-GR^2)}{U(n^>_1-\xi+\frac{1}{2},-\xi+\frac{3}{2},-GR^2)}=\frac{
\xi-\frac{1}{2}}{E+\tau \lam} \
\frac{M(n^<_1-\xi+\frac{1}{2},-\xi+\frac{1}{2},-GR^2)}{M(n^<_1-\xi+\frac{1}{2},-\xi+\frac{3}{2},-GR^2)}.
\eeq
It is important to note {that}
%equation~
\eqref{eqW.car1} is obtained {from}
%equation~
\eqref{eqS.car2} by the following mappings
\beq
 \xi \longrightarrow - \xi, \qquad \tau
\longrightarrow - \tau, \qquad G \longrightarrow - G
\eeq
which can also be used to derive
%Also, with the
%same map we {obtain}
%equation~
\eqref{eqW.car2} {from}
%equation~
\eqref{eqS.car1} in similar way as before.
Moreover, the bound-state levels for negative effective magnetic
field can be found by using the symmetry
\beq
E(\xi, \tau, G) =
E(-\xi, -\tau, -G).
\eeq
In both cases (positive and negative
effective magnetic fields), the valley degeneracy is controllably
broken by the presence of the effective magnetic field corresponding to our system.

%%%%%%%%%%%%%%%%%%%%%%%%%%%%%%%%%%%%%%%%%%%%%%
\section{Conclusion}
%%%%%%%%%%%%%%%%%%%%%%%%%%%%%%%%%%%%%%%%%%%%%%%

We considered the Dirac equation in (2+1)-dimensions in the presence of a constant magnetic field together
with a two-dimensional Dirac-oscillator potential coupling.
Indeed, using a similarity transformation, we formulated our
problem in terms of the polar coordinate representation that
allows us to handle easily the exact relationship between spinor
components. Then, we got the solutions to the energy spectrum of
the Dirac equation for a {gate-tunable} graphene
QD.

We discussed our results based on different physical
parameters. In fact, we considered three cases, whether the
cyclotron frequency is similar
or larger/smaller compared to the oscillator frequency.
%than $\om_c$.
This suggests defining an effective
magnetic field that produces the effective quantized Landau
levels. Our results were employed to discuss three important
limiting cases those concern the weak, strong and fine tuned magnetic
field situations.

We  studied the effective magnetic field
dependence of energy levels. Indeed, we  showed that the degeneracy of
the valley is controllably broken by the effective magnetic field.
In the fine tuned magnetic field cases (zero effective magnetic
field), we found that the degeneracy of the levels in the two
valleys is clearly displayed. Then, we  analyzed in detail the
case when the cyclotron frequency is much larger than the
oscillator frequency (strong magnetic field). It has also been
shown that at zero effective magnetic field, the valley degeneracy
are not broken. However, by increasing the effective field we  showed
 that the valley degeneracy are broken.

 In addition, when
the magnetic field is turned off the problem becomes a pure Dirac
oscillator, which leads to the creation of a negative effective
magnetic field. These results showed that it is possible to
control both spin and valley degeneracy by the effective magnetic
field corresponding to our system. One can note that our results have an importance in forming valley filters
 as well as
spin qubits in graphene quantum dots.

%Needless to say that one may obtain identical results for

\end{document}